\documentclass[12pt]{extarticle}
\usepackage{setspace}

\usepackage[T1]{fontenc}

\usepackage[latin1]{inputenc}
\usepackage{epsfig}
\usepackage[english,french]{babel}
\usepackage{color}
\usepackage{graphicx}

\usepackage{dcolumn}

\usepackage{moreverb}

\usepackage{amsmath,amssymb,amsfonts}

\begin{document}
\numberwithin{equation}{section}
\newcommand{\boxedeqn}[1]{%
  \[\fbox{%
      \addtolength{\linewidth}{-2\fboxsep}%
      \addtolength{\linewidth}{-2\fboxrule}%
      \begin{minipage}{\linewidth}%
      \begin{equation}#1\end{equation}%
      \end{minipage}%
    }\]%
}


\newsavebox{\fmbox}
\newenvironment{fmpage}[1]
     {\begin{lrbox}{\fmbox}\begin{minipage}{#1}}
     {\end{minipage}\end{lrbox}\fbox{\usebox{\fmbox}}}

\raggedbottom
\onecolumn

\begin{center}
\title*{{\LARGE{\textbf{Superintegrability and higher order polynomial algebras II}}}}
\end{center}
Ian Marquette
\newline
D\'epartement de physique et Centre de recherche math\'ematique,
Universit\'e de Montr\'eal,
\newline
C.P.6128, Succursale Centre-Ville, Montr\'eal, Qu\'ebec H3C 3J7,
Canada
\newline
ian.marquette@umontreal.ca
\newline
\newline
\newline
In an earlier article, we presented a method to obtain integrals of motion and polynomial algebras for a class of two-dimensional superintegrable systems from creation and annihilation operators. We discuss the general case and present its polynomial algebra. We will show how this polynomial algebra can be directly realized as a deformed oscillator algebra. This particular algebraic structure allows to find the unitary representations and the corresponding energy spectrum. We apply this construction to a family of caged anisotropic oscillators. The method can be used to generate new superintegrable systems with higher order integrals. We obtain new superintegrable systems involving the fourth Painlev\'e transcendent and present their integrals of motion and polynomial algebras. 
\newline
\section{Introduction}
The following article is the second of a series of two [1] discussing the construction of higher order integrals of motion and polynomial algebras from creation and annihilation operators. We applied the results to two cases with a second and a third order integrals where no polynomial algebra were found
\newline
\newline
\begin{equation}
V(x,y)=\hbar^{2}[\frac{1}{8a^{4}}(x^{2}+y^{2})+\frac{1}{y^{2}}+\frac{1}{(x+a)^{2}}+\frac{1}{(x-a)^{2}}
] ,
\end{equation}
\begin{equation}
V(x,y)=\hbar^{2}[\frac{1}{8a^{4}}(x^{2}+y^{2})+\frac{1}{(y+a)^{2}}+\frac{1}{(y-a)^{2}}.
+\frac{1}{(x+a)^{2}}+\frac{1}{(x-a)^{2}} ] .
\end{equation}
\newline
\newline
We constructed a quintic and a seventh order algebras. We studied the realization in terms of deformed oscillator algebras of a class of polynomial algebras of the seventh order. These results allowed us to obtain the structure function for the Potential 5 and 6 and unitary representations with the corresponding energy spectrum.
\newline
Many article [3-18] were devoted to superintegrable systems. However, most articles were on quadratically superintegrable systems. This paper follows also articles concerning superintegrable systems with third order integrals of motion [19,20,21,22,23,24]. The potentials given by Eq.(1.1) and (1.2) were obtained in Ref. 20 and studied from the point of view of supersymmetric quantum mechanics in Ref. 23. In Ref. 20 a potential written in terms of the fourth Painlev\'e transcendent was found. 
\newline
\begin{equation}
V(x,y)=\frac{\omega^{2}}{2}(x^{2}+y^{2})+\epsilon\frac{\hbar\omega}{2}f^{'}(\sqrt{\frac{\omega}{\hbar}}x)+\frac{\omega\hbar}{2}f^{2}(\sqrt{\frac{\omega}{\hbar}}x)
\end{equation}
\[+\omega \sqrt{\hbar \omega}xf(\sqrt{\frac{\omega}{\hbar}}x)+\frac{\hbar\omega}{3}(-\alpha+\epsilon) \quad . \]    
We will use the potential given by Eq.(1.3) to generate new superintegrable systems with a higher order integrals of motion. By construction, they also have a second order integral of motion. This integral is related to separation of variables.
\newline
\newline
Let us present the organization of this paper. In Section 2, we recall how we can generate integrals of motion from the creation and annihilation operators. We consider the general case $m\lambda_{x}=n\lambda_{y}$ and obtain the polynomial algebra generated by integrals of motion. We show how we can obtain directly the realization in terms of deformed oscillator algebras. In Section 3, we apply the construction to a family of caged anisotropic harmonic oscillator. We show how the method can be used to generate new superintegrable systems with higher order integrals. We construct a new family of Hamiltonians written in terms of the fourth Painlev\'e transcendent. We present their integrals, polynomial algebra and energy spectrum. These results extend the number of known superintegrable systems involving the Painlev\'e transcendents
\newline
\section{Polynomial algebras}
Let us consider a two-dimension Hamiltonian separable in Cartesian coordinates 
\begin{equation}
H(x,y,P_{x},P_{y})=H_{x}(x,P_{x})+H_{y}(y,P_{y}),
\end{equation}
\newline
for which creation and annihilation operators ( polynomials in momenta) $A_{x}$, $A_{x}^{\dagger}$, $A_{y}$ and $A_{y}^{\dagger}$ exist. These operators satisfy
\newline
\begin{equation}
[H_{1},A_{x}^{\dagger}]=\lambda_{x}A_{x}^{\dagger},\quad [H_{2},A_{y}^{\dagger}]=\lambda_{y}A_{y}^{\dagger}\quad .
\end{equation}
The following operators
\begin{equation}
f_{1}=A_{x}^{\dagger m}A_{y}^{n},\quad f_{2}=A_{x}^{m}A_{y}^{\dagger n}\quad ,
\end{equation}
commute with the Hamiltonian H
\begin{equation}
[H,f_{1}]=[H,f_{2}]=0,
\end{equation}
if 
\begin{equation}
m\lambda_{x}-n\lambda_{y}=0, \quad  m,n \in \mathbb{Z}^{+} \quad .
\end{equation}
\newline
Creation and annihilation operators allow to construct polynomial integrals of motion.
\newline
We will now consider integrals given by Eq.(2.7) and show that we can obtain directly a polynomial algebra written as a deformed oscillator algebra. Let us consider
\begin{equation}
H=H_{x}+H_{y}, \quad \lambda=m\lambda_{x}=n\lambda_{y},
\end{equation}
\begin{equation}
A=\frac{1}{2\lambda}(H_{x}-H_{y}),\quad I_{-}=A_{x}^{m}A_{y}^{\dagger n},\quad I_{+}=A_{x}^{\dagger m}A_{y}^{n}
\end{equation}
\newline
We demand that creation and annihilation operators satisfy the following relations
\newline
\begin{equation}
[A_{x},A_{x}^{\dagger}]=P(H_{x})=Q(H_{x}+\lambda_{x})-Q(H_{x}),
\end{equation}
\begin{equation}
[A_{y},A_{y}^{\dagger}]=R(H_{y})=S(H_{y}+\lambda_{y})-S(H_{y}).
\end{equation}
\newline
We obtain the polynomial algebra
\newline
\begin{equation}
[A,I_{-}]=-I_{-},\quad [A,I_{+}]=I_{+},
\end{equation}
\begin{equation}
[I_{-},I_{+}]=\prod_{l=1}^{m}Q(\frac{H}{2}+m\lambda_{x}A+l\lambda_{x})\prod_{k}^{n}S(\frac{H}{2}-n\lambda_{y}A-(n-k)\lambda_{y})
\end{equation}
\[-\prod_{i=1}^{m}Q(\frac{H}{2}+m\lambda_{x}A-(m-i)\lambda_{x})\prod_{j}^{n}S(\frac{H}{2}-n\lambda_{y}A+j\lambda_{y})\]
\newline
The order of the polynomial algebra is thus determine by the order of polynomials Q and S. This last relation have a very particular structure and we rewrite the Eq.(2.11) as
\newline
\begin{equation}
[I_{-},I_{+}]=F_{m,n}(H,A+1)-F_{m,n}(H,A),
\end{equation}
with
\begin{equation}
F_{m,n}=\prod_{i=1}^{m}Q(\frac{H}{2}+m\lambda_{x}A-(m-i)\lambda_{x})\prod_{j}^{n}S(\frac{H}{2}-n\lambda_{y}A+j\lambda_{y})
\end{equation}
\newline
We can define
\begin{equation}
b^{\dagger}=I_{+},\quad b=I_{-},\quad N=A-u
\end{equation}
\begin{equation}
[N,b^{\dagger}]=b^{\dagger},\quad [N,b]=-b,\quad b^{\dagger}b=\Phi(H,N),\quad bb^{\dagger}=\Phi(H,N+1)
\end{equation}
and
\begin{equation}
\Phi(H,N)=F_{m,n}(H,N+u)
\end{equation}
\newline
To obtain Fock type unitary representations and their corresponding energy spectrum we impose the following conditions
\newline
\begin{equation}
\Phi(p+1,u_{i},k)=0, \quad \Phi(0,u,k)=0,\quad \phi(x)>0, \quad \forall \quad x>0 \quad .
\end{equation}
\newline
We showed that the general polynomial algebra in the x and y axis given by Eq(2.2), (2.8) and (2.9) with $m\lambda_{x}=n\lambda_{y}$ allow us to generate a higher order polynomial algebra for the superintegrable system given by Eq.(2.10) and (2.11). This polynomial algebra is directly written as a deformed oscillator algebra. 
\newline 
The two-dimensional anisotropic oscillator is a particular case and was studied in Ref. 25. From the point of view of physics this system is important in nuclear and atomic physics. In nuclear physics it describes single-particle level spectrum of pancake i.e. triaxially deformed nuclei with $\omega_{x} >>\omega_{y},\omega_{z}$. The system with ratio 2:1 describes superdeformed nuclei and ratio 3:1 hyperdeformed nuclei. We will consider generalization of this systems that could have application in nuclear physics.
\section{Applications}
\subsection{Caged anisotropic harmonic oscillator}
The anisotropic harmonic oscillator can be generalized by adding singular terms. This system is the caged anisotropic harmonic oscillator [26,27]
\newline
\begin{equation}
H=\frac{P_{x}^{2}}{2}+\frac{P_{y}^{2}}{2}+\frac{\omega^{2}}{2}(k^{2}x^{2}+m^{2}y^{2})+\frac{l_{1}}{x^{2}}+\frac{l_{2}}{y^{2}}.
\end{equation}
\newline
The method of separation of variables allows to solve the corresponding Schrödinger equation in terms of Laguerre polynomials [8] and to obtain the energy spectrum. However, the polynomial algebra remains to be determined. We apply to this system the construction of Section 2.
\newline
\newline
We have the following operators
\newline
\begin{equation}
A_{x}^{\dagger}=-\frac{1}{4}(\frac{\hbar}{\omega k}\frac{d^{2}}{dx^{2}}-2x\frac{d}{dx}+\frac{\omega k}{\hbar}x^{2}-\frac{2l_{1}}{\omega k \hbar x^{2}}-1),
\end{equation}
\[ A_{x}=-\frac{1}{4}(\frac{\hbar}{\omega k}\frac{d^{2}}{dx^{2}}+2x\frac{d}{dx}+\frac{\omega k}{\hbar}x^{2}-\frac{2l_{1}}{\omega k \hbar x^{2}}+1),\]
\[ A_{y}=-\frac{1}{4}(\frac{\hbar}{\omega m}\frac{d^{2}}{dy^{2}}-2y\frac{d}{dy}+\frac{\omega m}{\hbar}y^{2}-\frac{2l_{2}}{\omega m \hbar y^{2}}-1),\]
\[ A_{y}^{\dagger}=-\frac{1}{4}(\frac{\hbar}{\omega m}\frac{d^{2}}{dy^{2}}+2y\frac{d}{dy}+\frac{\omega m}{\hbar}y^{2}-\frac{2l_{2}}{\omega m \hbar y^{2}}+1).\]
\newline
They satisfy the relations given by Eq.(2.2) and (2.5) with $\lambda_{x}=2\hbar k \omega$, $\lambda_{y}=2\hbar m \omega$ and
$m\lambda_{x}=k \lambda_{y}$. We can apply results of Section 2.  We have
\newline
\begin{equation}
Q(H_{x})=\frac{1}{4\hbar^{2}k^{2}\omega^{2}}H_{x}^{2}-\frac{1}{2\hbar k\omega}H_{x}+(\frac{3}{16}-\frac{l_{1}}{2\hbar^{2}}),
\end{equation}
\begin{equation}
S(H_{y})=\frac{1}{4\hbar^{2}m^{2}\omega^{2}}H_{y}^{2}-\frac{1}{2\hbar m\omega}H_{y}+(\frac{3}{16}-\frac{l_{2}}{2\hbar^{2}}).
\end{equation}
The integrals are given by Eq.(2.7). We have the condition $m,k \in \mathbb{Z}^{+}$. The Eq.(3.6) gives 
\newline
\begin{equation}
\Phi_{m,k}(x)=m^{2}k^{2}\prod_{i=1}^{m}(\frac{E}{4mk\hbar\omega}+x+u-1+\frac{i}{m}-\frac{1}{2m}-\frac{\nu_{1}}{2m})
\end{equation}
\[ (\frac{E}{4mk\hbar\omega}+x+u-1+\frac{i}{m}-\frac{1}{2m}+\frac{\nu_{1}}{2m})\]
\[\prod_{j=1}^{k}(\frac{E}{4mk\hbar\omega}-x-u+\frac{j}{k}-\frac{1}{2k}-\frac{\nu_{2}}{2k})\]
\[ (\frac{E}{4mk\hbar\omega}-x-u+\frac{j}{k}-\frac{1}{2k}+\frac{\nu_{2}}{2k}),\]
with
\[\nu_{1}=\sqrt{1+\frac{8l_{1}}{\hbar^{2}}},\quad \nu_{2}=\sqrt{1+\frac{8l_{2}}{\hbar^{2}}}.\]
\newline
We should impose the constraints given by the Eq.(2.17). We obtain the following solutions
\newline
\begin{equation}
u=\frac{-E}{4mk\hbar\omega}+\frac{m-p}{m}+\frac{1}{2m}+\frac{\epsilon_{1}\nu_{1}}{2m},
\end{equation}
\begin{equation}
\Phi_{m,k}(x)=m^{2}k^{2}\prod_{i=1}^{m}(x+\frac{i-p}{m})(x+\frac{i-p}{m}+\frac{\epsilon_{1}\nu_{1}}{m})
\end{equation}
\[ \prod_{j=1}^{k}(N+1+\frac{j-q}{k}-x)(N+1+\frac{j-q}{k}+\frac{\epsilon_{2}\nu_{2}}{k}),\]
\newline
\begin{equation}
E=2mk\hbar\omega(N+2+\frac{1-2p+\epsilon_{1}\nu_{1}}{2m}+\frac{1-2q+\epsilon_{2}\nu_{2}}{2k}).
\end{equation}
$p=1,2,...,m$, $q=1,2,...,k$ and $N \in \mathbb{N}$.
\subsection{System with Painlev\'e transcendent} 
In Ref. 20, five systems involving Painlev\'e transcendent [28] were found. One of these systems was written as a function of the fourth Painlev\'e transcendent. 
\newline
\begin{equation}
V(x,y)=\frac{\omega^{2}}{2}(x^{2}+y^{2})+\epsilon\frac{\hbar\omega}{2}f^{'}(\sqrt{\frac{\omega}{\hbar}}x)+\frac{\omega\hbar}{2}f^{2}(\sqrt{\frac{\omega}{\hbar}}x)
\end{equation}
\[+\omega \sqrt{\hbar \omega}xf(\sqrt{\frac{\omega}{\hbar}}x)+\frac{\hbar\omega}{3}(-\alpha+\epsilon) \quad . \]           
\newline
We presented its cubic algebra, wave functions and creation and annihilation operators. The Hamiltonians with $\epsilon=1$ and $\epsilon=-1$ are related by a special case of third order supersymmetry called shape invariance [24,29].
\newline
Let us consider the following superintegrable system
\newline
\begin{equation}
H=\frac{P_{x}^{2}}{2}+\frac{P_{y}^{2}}{2}+g_{1}(x)+g_{2}(y) \quad ,
\end{equation}
\begin{equation}
g_{1}(x)=\frac{\omega_{1}^{2}}{2}x^{2}+\frac{\hbar\omega_{1}\epsilon_{1}}{2}f_{1}^{'}(\sqrt{\frac{\omega_{1}}{\hbar}}x)+\frac{\omega_{1}\hbar}{2}f_{1}^{2}(\sqrt{\frac{\omega_{1}}{\hbar}}x)+\omega_{1} \sqrt{\hbar \omega_{1}}xf_{1}(\sqrt{\frac{\omega_{1}}{\hbar}}x)+\frac{\hbar\omega_{1}}{3}(-\alpha_{1}+\epsilon_{1}) \quad ,            
\end{equation}
\begin{equation}
g_{2}(y)=\frac{\omega_{2}^{2}}{2}y^{2}+\frac{\hbar\omega_{2}\epsilon_{2}}{2}f_{2}^{'}(\sqrt{\frac{\omega_{2}}{\hbar}}y)+\frac{\omega_{2}\hbar}{2}f_{2}^{2}(\sqrt{\frac{\omega_{2}}{\hbar}}y)+\omega_{2} \sqrt{\hbar \omega_{2}}yf_{2}(\sqrt{\frac{\omega_{2}}{\hbar}}y)+\frac{\hbar\omega_{2}}{3}(-\alpha_{2}+\epsilon_{2}) \quad ,            
\end{equation}
with 
\begin{equation}
m\omega_{1}=n\omega_{2},\quad \lambda_{x}=\hbar\omega_{1}=\tilde{\omega},\quad \lambda_{y}=\hbar\omega_{2}
\end{equation}
\newline
The function $f_{1}=f_{1}(x,\alpha_{1},\beta_{1})$ and $f_{2}=f_{1}(y,\alpha_{2},\beta_{2})$ are the fourth Painlev\'e transcendent.
The third order creation and annihilation operators were discussed in Ref.24 and Ref.29. The commutator of these operators was also obtained. They satisfy the relation given by Eq(2.8) and (2.9) with
\begin{equation}
Q(H_{x})=8(H_{x}-\frac{\hbar\omega}{3}(-\alpha_{1}+\epsilon_{1}+3))((H_{x}-\frac{\hbar\omega}{3}(\frac{\alpha_{1}}{2}+4\epsilon_{1}-\frac{3}{2})   )^{2}+ \frac{\omega^{2}\hbar^{2}\beta_{1}}{8})
\end{equation}
\begin{equation}
S(H_{y})=8(H_{y}-\frac{\hbar\omega}{3}(-\alpha_{2}+\epsilon_{2}+3))((H_{y}-\frac{\hbar\omega}{3}(\frac{\alpha_{2}}{2}+4\epsilon_{2}-\frac{3}{2}) )^{2}+ \frac{\omega^{2}\hbar^{2}\beta_{2}}{8})
\end{equation}
\newline
The structure function of the general case is given by Eq.(2.16)
\newline
\begin{equation}
\Phi_{m,n}(x)=\prod_{i=1}^{m}(\frac{E}{2}+m\hbar\omega_{1}(x+u)-(m-i)\hbar\omega_{1}-\gamma_{1}m\omega_{1}\hbar)
\end{equation}
\[ ( \frac{E}{2}+m\hbar\omega_{1}(x+u)-(m-i)\hbar\omega_{1}-\gamma_{2}m\omega_{1}\hbar  ) ( \frac{E}{2}+m\hbar\omega_{1}(x+u)-(m-i)\hbar\omega_{1} -\gamma_{3}m \omega_{1}\hbar  ) \]
\[\prod_{j=1}^{n}(\frac{E}{2}-n\hbar\omega_{2}(x+u)+j\hbar\omega_{2}-\gamma_{4}n\omega_{2}\hbar) (\frac{E}{2}-n\hbar\omega_{2}(x+u)+j\hbar\omega_{2} -\gamma_{5} n \omega_{2}\hbar  ) \]
\[  ( \frac{E}{2}-n\hbar\omega_{2}(x+u)+j\hbar\omega_{2} -\gamma_{6}n \omega_{2}\hbar ),\]
with
\begin{equation}
\gamma_{1}=-\frac{1}{3m}(-3+\alpha_{1}-\epsilon_{1}),
\gamma_{2}=\frac{\hbar\omega_{1}}{12m}(-6+2\alpha_{1}-3i\sqrt{2\beta_{1}}+16\epsilon_{1} ) ,
\end{equation}
\begin{equation}
\gamma_{3}=\frac{\hbar\omega_{1}}{12m}(-6+2\alpha_{1}+3i\sqrt{2\beta_{1}}+16\epsilon_{1} ) ,
\gamma_{4}=\frac{1}{3n}(-3+\alpha_{2}-\epsilon_{2}),
\end{equation}
\begin{equation}
\gamma_{5}=\frac{\hbar\omega_{1}}{12n}(-6+2\alpha_{2}-3i\sqrt{2\beta_{2}}+16\epsilon_{2} ) ,
\gamma_{6}=\frac{\hbar\omega_{1}}{12n}(-6+2\alpha_{2}+3i\sqrt{2\beta_{2}}+16\epsilon_{2} ) .
\end{equation}
\newline
We can rewrite the structure function as
\newline
\begin{equation}
\Phi_{m,n}(x)=\prod_{i=1}^{m}\tilde{\omega^{6}}\hbar^{6}(\frac{E}{2\hbar\tilde{\omega}}+x+u-1+\frac{i}{m}-\gamma_{1})
\end{equation}
\[ ( \frac{E}{2\hbar\tilde{\omega}}+x+u-1+\frac{i}{m}-\gamma_{2}  ) ( \frac{E}{2\hbar\tilde{\omega}}+x+u-1+\frac{i}{m} -\gamma_{3}  ) \]
\[\prod_{j=1}^{n}(\frac{E}{2\hbar\tilde{\omega}}-x-u+\frac{j}{n}-\gamma_{4}) (\frac{E}{2\hbar\tilde{\omega}}-x-u+\frac{j}{m} -\gamma_{5}    ) \]
\[  ( \frac{E}{2\hbar\tilde{\omega}}-x-u+\frac{j}{n} -\gamma_{6}  ).\]
\newline
We obtain the finite dimensional unitary representations and the corresponding energy spectrum from the Eq.(4.7). We have
\newline
\begin{equation}
u_{1}=\frac{-E}{2\hbar\tilde{\omega}}+1-\frac{p}{m}+\gamma_{1},u_{2}=\frac{-E}{2\hbar\tilde{\omega}}+1-\frac{p}{m}+\gamma_{2},
u_{3}=\frac{-E}{2\hbar\tilde{\omega}}+1-\frac{p}{m}+\gamma_{3}.
\end{equation}
Let us present one of the three solutions for $u_{1}$ 
\newline
\begin{equation}
E=\hbar\tilde{\omega}(N+2-\frac{p}{m}-\frac{q}{n}+\gamma_{1}+\gamma_{4}),
\end{equation}
\begin{equation}
\Phi_{1}=\prod_{i}^{m}\prod_{j}^{n}\tilde{\omega^{6}}\hbar^{6}(x+\frac{i-p}{m})(x+\frac{i-p}{m}+\gamma_{1}-\gamma_{2})
\end{equation}
\[(x+\frac{i-p}{m}+\gamma_{1}-\gamma_{3})(N+1-x+\frac{j-q}{n})\]
\[ (N+1-x+\frac{j-q}{n}+\gamma_{4}-\gamma_{5})(N+1-x+\frac{j-q}{n}+\gamma_{4}-\gamma_{6}).\]
$p=1,2,...,m$, $q=1,2,...,k$ and $N$ $\in$ $\mathbb{N}$.
\newline
There are 6 other solutions of the same form for $u_{2}$ and $u_{3}$. 
\section{Conclusion}
The main result of this article is that we constructed in Section 2 the polynomial algebra for the general case $m \lambda=n\lambda$ and showed that this algebra can be realized directly as a deformed oscillator algebra. This result allows to study many systems that could have applications in nuclear physics, atomic physics and quantum chemistry. 
\newline
We showed how the construction can be used to generate new superintegrable systems from known one-dimensional Hamiltonians with creation and annihilation operators. This result is also valid in classical mechanics.
\newline
We studied with this method a family of caged anisotropic oscillator. We found the polynomial algebra, the finite dimensional unitary representations and the energy spectrum. We applied also the method to a new superintegrable system involving the fourth Painlev\'e transcendent. We found the polynomial algebra, the finite dimensional unitary representations and the corresponding energy spectrum. The two systems given by Eq.(3.11), (3.10), (3.11) and (3.12) can be generalized in higher dimensions.
\newline
The classification of systems with creation and annhilation operators is important and could allow to find new superintegrable systems. A classification of second order creation and annihilation operators was discussed in Ref. 30. To our knowledge the classification of systems with second order ladder operators is not complete and only a class of Hamiltonians with third order ladder operators were discussed [29,24]. 
\newline
\newline
\textbf{Acknowledgments} The research of I.M. was supported by a postdoctoral
research fellowship from FQRNT of Quebec. The author thanks P.Winternitz for very helpful comments and discussions.

\section{\textbf{References}}

1. I.Marquette, Superintegrability and higher order polynomial algebras I (2009).
\newline
2. V.A.Dulock and H.V.McIntosh, Am. J.Phys. 33, 109 (1965).
\newline
3. V.Fock, Z.Phys. 98, 145-154 (1935).
\newline
4. V.Bargmann, Z.Phys. 99, 576-582 (1936).
\newline
5. J.M.Jauch and E.L.Hill, Phys.Rev. 57, 641-645 (1940).
\newline
6. M.Moshinsky and Yu.F.Smirnov, The Harmonic Oscillator In Modern
Physics, (Harwood, Amsterdam, 1966).
\newline
7. J.Fris, V.Mandrosov, Ya.A.Smorodinsky, M.Uhlir and P.Winternitz, Phys.Lett. 16, 354-356 (1965).
\newline
8. P.Winternitz, Ya.A.Smorodinsky, M.Uhlir and I.Fris, Yad.Fiz. 4,
625-635 (1966). (English translation in Sov. J.Nucl.Phys. 4,
444-450 (1967)).
\newline
9. A.Makarov, Kh. Valiev, Ya.A.Smorodinsky and P.Winternitz, Nuovo Cim. A52, 1061-1084 (1967).
\newline
10. N.W.Evans, Phys.Rev. A41, 5666-5676 (1990), J.Math.Phys. 32,
3369-3375 (1991).
\newline
11. E.G.Kalnins, J.M.Kress, W.Miller Jr and P.Winternitz,
J.Math.Phys. 44(12) 5811-5848 (2003).
\newline
12. E.G.Kalnins, W.Miller Jr and G.S.Pogosyan, J.Math.Phys. A34,
4705-4720 (2001).
\newline
13. E.G.Kalnins, J.M.Kress and W.Miller Jr, J.Math.Phys. 46,
053509 (2005), 46, 053510 (2005), 46, 103507 (2005), 47, 043514
(2006), 47, 043514 (2006).
\newline
14. E.G.Kalnins, W.Miller Jr and G.S.Pogosyan, J.Math.Phys. 47,
033502.1-30 (2006), 48, 023503.1-20 (2007).
\newline
15. J.Daboul, P.Slodowy et C.Daboul, Phys.Lett. B317, 321-328 (1993)
\newline
16. Ya.I.Granovsky, A.S.Zhedanov et I.M.Lutzenko, Ann. Phys. (New York) 217, 1, 1-20 (1992).
\newline
17. P.L\'etourneau et L.Vinet, Ann. Phys. (New York) 243, 144-168 (1995).
\newline
18. C.Daskaloyannis, J.Math. Phys. 42, 1100-1119 (2001).
\newline
19. S.Gravel and P.Winternitz, J.Math.Phys. 43(12), 5902 (2002).
\newline
20. S.Gravel, J.Math.Phys. 45(3), 1003-1019 (2004).
\newline
21. I.Marquette and P.Winternitz, J.Math.Phys. 48(1) 012902
(2007).
\newline
22. I.Marquette and P.Winternitz, J. Phys. A: Math. Theor. 41, 304031 (2008).
\newline
23. I.Marquette, J. Math. Phys. 50, 012101 (2009).
\newline
24.  I.Marquette, J.Math.Phys. 50 095202 (2009).
\newline
25. D.Bonatsos, C.Daskaloyannis, Prog.Part.Nucl.Phys. 43, 537 (1999).
\newline
26. P.E.Verrier and N.W.Evans, J.Math.Phys. 49, 022902 (2008).
\newline
27. M.A.Rodriguez, P.Tempesta and P.Wintertnitz, Phys. Rev. E78, 046608 (2008).
\newline
28. E.L.Ince, Ordynary Differential Equations, Dover, New York (1956).
\newline
29. A.Andrianov, F.Cannata, M.Ioffe and D.Nishnianidze, Phys.Lett.A, 266,341-349 (2000).
\newline
30. C.P.Boyer and W.Miller Jr., J.Math.Phys. 15, 9 (1974).

\end{document}